\documentstyle[preprint,aps]{revtex}

\def\beq{\begin{equation}}
\def\eeq{\end{equation}}

\input amssym.def

\makeatletter
\def\theequation{\arabic{section}.\theequation@prefix\arabic{equation}}%
\def\mathletters{%
\inc@eqnnum  \setcounter{eqletter}{0}%
\edef\@currentlabel{\theequation}%
\def\theequation{\arabic{section}.\theequation@prefix\arabic{equation}%
           \alph{eqletter}}%
\def\inc@eqnnum{\addtocounter{eqletter}{1}}%
\def\dec@eqnnum{\addtocounter{eqletter}{-1}}%
}%
\def\section{{\setcounter{equation}{0}}
\@mainheadtrue
\@startsection {section}{1}{\z@}{0.8cm plus1ex minus
 .2ex}{0.5cm plus1ex minus.2ex}{\reset@font\small\bf\centering}
           }
\makeatother

\begin{document}

\def\footnoterule{\hrule width \hsize}
\skip\footins = 12pt
\footskip     = 18pt
\footnotesep  = 10pt


\textwidth=6.5in
\hsize=6.5in
\oddsidemargin=0in
\evensidemargin=0in
\hoffset=0in

\textheight=9.5in
\vsize=9.5in
\topmargin=-.5in
\voffset=-.3in

\title{Energy-Momentum Tensor Improvements in Two Dimensions\footnotemark[1]}

\def\footstrut{\baselineskip 16pt}

\footnotetext[1]
{\baselineskip=16pt
This work is supported in part
by funds provided by
NSF grant PHY--9315811 and by
the U.S.~Department of Energy (D.O.E.)
under contract \#DE-FC02-94ER40818.
\hfil\break
Brandeis BRX-TH-384, MIT-CTP-2469 \hfil
October 1995\break}

\author{S.~Deser}
\medskip
\address{Department of Physics, Brandeis University,
Waltham, MA ~02254--9110}

\bigskip

\author{R. Jackiw}

\address{Center for Theoretical Physics, Laboratory for Nuclear Science
and Department of Physics \\[-1ex]
Massachusetts Institute of Technology, Cambridge, MA ~02139--4307}

\maketitle

\setcounter{page}{0}
\thispagestyle{empty}

\begin{abstract}%
\noindent%
\baselineskip=16pt%

We discuss some aspects of the two-dimensional scalar field, considering
particularly the action for the conformal anomaly as an
``improved'' gravitational coupling, and the possibility of introducing a dual
coupling, which provides a ``chiral'' energy-momentum tensor improvement.

\end{abstract}

\vspace*{\fill}
\begin{center}
{\it In Memoriam:} H.~Umezawa
\end{center}
\vspace*{\fill}

\newpage

\widetext

There is very little that is not known about scalar fields and anomalies
of conformal (Weyl) symmetry in two spacetime dimensions.  In this note,
dedicated to the memory of H.~Umezawa, who contributed to our
understanding of symmetry behavior in field theories, we discuss some less
emphasized aspects of the subject, although our remarks may not surprise
experts \cite{ref:1}.  We shall be especially concerned with the
interplay of ``energy-momentum tensor improvements'' and Schwinger terms
/ Virasoro algebra central charges, as well as novel, chiral improvements.

\setcounter{section}{1}

{\bf I}.~
Consider first the action of a scalar field,
coupled non-minimally to gravity
\beq
I = -{1\over48\pi} \int d^2 x \, \sqrt{-g}
\left( {1\over2} g^{\mu\nu} \partial_\mu \varphi \partial_\nu \varphi
+ R \varphi \right) \: .
\label{eq:1.1}
\eeq
[Our conventions are: signature $(+,-)$,
$R_{\mu\nu} = - \partial_\alpha \Gamma^\alpha_{\mu\nu} + \ldots$,
$x^\mu = (t,\sigma)$.]
This ``improved'' scalar field action is of course also the local
version of the Polyakov action $I_P$.
\begin{mathletters}%
\beq
I_P = {1\over96\pi} \int d^2 x \, d^2 y \sqrt{-g(x)}
R(x) K^{-1}(x,y) \sqrt{-g(y)} R(y)
\label{eq:1.2a}
\eeq
\beq
- \sqrt{-g(x)} \, D_x^2 \,
K^{-1} (x,y) =
\delta^2 (x-y)
\label{eq:1.2b} 
\eeq
\end{mathletters}%
%
%
Eq.~(\ref{eq:1.2a}) is obtained from (\ref{eq:1.1})
upon eliminating $\varphi$
through the usual Gaussian shift,
and the field $\varphi$ obeys
\beq
D^2 \varphi = R \: .
\label{eq:1.3}
\eeq
Here $D^2 = g^{\mu\nu} D_\mu D_\nu$
is the covariant d'Alembertian acting on scalars
and the $\delta$-function is a density.
The normalization of the
integral in (\ref{eq:1.1}) is chosen to agree with
(\ref{eq:1.2a}).
[In fact one may set the $R\varphi$ coefficient to be an arbitrary
number $n$, provided the normalizing factor is divided by $n^2$; this
does not change the overall negative sign in (\ref{eq:1.1}).]

The metric variation of $I_P$
defines the energy-momentum tensor $\theta_{\mu\nu}$,
\beq
{2\over\sqrt{-g}} \,
{\delta I_P \over \delta g^{\mu\nu}}
\equiv
\theta_{\mu\nu}
\label{eq:1.4}
\eeq
\begin{mathletters}%
\label{eq:1.5}
\beq
\theta_{\mu\nu} = - {1\over 48\pi}
\left( \partial_\mu \phi \partial_\nu \phi
- {1\over2} g_{\mu\nu} g^{\alpha\beta} \partial_\alpha \phi
\partial_\beta \phi
\right)
- {1\over 24\pi}
\left( D_\mu D_\nu - g_{\mu\nu} D^2 \right) \phi
\label{eq:1.5a}
\eeq
where $\phi$ is a solution of
\beq
D^2 \phi = R \: .
\label{eq:1.5b}
\eeq
\end{mathletters}%
As is well known, $I_P$ emerges when a massless, scalar quantum field
$\Phi$, minimally coupled to the metric tensor $g_{\mu\nu}$, is
integrated in the functional integral for the partition function
\cite{ref:2}.
In this framework, $\theta_{\mu\nu}$ may be viewed
as the vacuum expectation
value of the $\Phi$-field's
energy-momentum tensor operator
\beq
\Theta_{\mu\nu} = \partial_\mu \Phi \partial_\nu \Phi
- {\textstyle{1\over2}} g_{\mu\nu} g^{\alpha\beta} \partial_\alpha \Phi
\partial_\beta \Phi
\label{eq:1.6}
\eeq
in the presence of external gravity \cite{ref:3}
\beq
\theta_{\mu\nu} =
{}_g \! \left\langle \Omega | \Theta_{\mu\nu} | \Omega \right\rangle \!
{}_g \: .
\label{eq:1.7}
\eeq

The evaluation of the partition function leading to
(\ref{eq:1.2a})
or of the matrix element (\ref{eq:1.7})
requires
a choice of regularization.
The exhibited formulas reflect the
option of maintaining
diffeomorphism invariance:
under an infinitesmal coordinate transformation,
\beq
\delta_f x^\mu = -f^\mu (x)
\label{eq:1.8}
\eeq
which is implemented on $g_{\mu\nu}$ by the Lie derivative $L_f$
with respect to the vector $f^\mu$,
\beq
\delta_f g_{\mu\nu} = L_f g_{\mu\nu}
\label{eq:1.9}
\eeq
$I_P$ in Eq.~(\ref{eq:1.2a}) is invariant and
(\ref{eq:1.6}) -- (\ref{eq:1.7}) are covariant.  However, Weyl
invariance is lost: the formal tracelessness of (\ref{eq:1.6})
is replaced by the anomaly equation, which follows from (\ref{eq:1.5})
\cite{ref:4}
\beq
\theta^\mu_\mu = {1\over 24\pi} R \: .
\label{eq:1.10}
\eeq

This same result may be gotten from the action (\ref{eq:1.1}), when its
fields are subjected to a combined Weyl variation
\begin{mathletters}%
\label{eq:1.11}
\begin{eqnarray}
\delta g_{\mu\nu} &=& \chi \, g_{\mu\nu}
\label{eq:1.11a} \\
\delta \varphi &=& \chi
\label{eq:1.11b}
\end{eqnarray}
\end{mathletters}%
\beq
2 \left. {\delta I \over \delta \chi} \right|_{\chi=0}
= -{1\over 24\pi} \sqrt{-g} R \: .
\label{eq:1.12}
\eeq
Therefore, it is recognized that (\ref{eq:1.1})  is an effective action
for the Weyl anomaly --- in that sense it is a Wess-Zumino (WZ)
-- like
anomaly action.   [Our conventions differ from those used in WZ
treatments, where $\delta g_{\mu\nu} = 2 \chi g_{\mu\nu}$,
$\delta \varphi = \chi$; this accounts for the relative ${1\over2}$
factor between the two terms in (\ref{eq:1.1}).  Our scaling
(\ref{eq:1.11}) is of course consistent with (\ref{eq:1.3}).]

An alternative way to see the connection between the
nonlocal but
purely metric action $I_P[g]$ and its local but $\varphi$-dependent WZ
formulation (\ref{eq:1.1}) is the following two-dimensional version of a
general construction \cite{ref:5}.  Consider the difference
\beq
\bar{I} \equiv I_P [g] - I_P [\tilde{g}] ~~, ~~~~
\tilde{g}_{\mu\nu} \equiv g_{\mu\nu} e^{-\varphi} ~~.
\label{eq:1.13}
\eeq
Since the combination $\tilde{g}$ is Weyl-invariant by (\ref{eq:1.11}),
it is obvious that $\bar{I}$ yields the same anomaly as $I_P[g]$.
If one now rewrites $I_P[\tilde{g}]$ explicitly using the identity
\beq
\sqrt{-\tilde{g}} \, R(\tilde{g}) = \sqrt{-g} \, R(g) -
\sqrt{-g} D^2 \varphi
\label{eq:1.14}
\eeq
and the fact that $K^{-1}$ is insensitive to conformal redefinition of
the metric,
then (\ref{eq:1.13}) is precisely (\ref{eq:1.1}).  In principle, one
could add any other Weyl-invariant
term to (\ref{eq:1.13}); while this ambiguity is
irrelevant here, it is in fact important in higher dimensions~\cite{ref:5}.

\def\ignore{
The above is the conventional description.  However, an alternative
formulation is also available.  Using well-known rules for transforming
$R$ by scaling $g_{\mu\nu}$, we may rewrite (\ref{eq:1.1}) as
\beq
I = -{1\over 48\pi} \int \sqrt{-\tilde{g}} \, R(\tilde{g}) \varphi
\label{eq:1.13}
\eeq
where
\beq
\tilde{g}_{\mu\nu} = e^{-\varphi/2} g_{\mu\nu}
\label{eq:1.14}
\eeq
Moreover, $\tilde{g}_{\mu\nu}$ is invariant against Weyl rescaling,
provided the transformation rule for $\varphi$ is changed from
(\ref{eq:1.11b}).
\begin{mathletters}%
\label{eq:1.15}
\begin{eqnarray}
\delta \, g_{\mu\nu} &=& \chi \, g_{\mu\nu} \label{eq:1.15a} \\
\delta \varphi &=& 2 \chi \label{eq:1.15b}
\end{eqnarray}
\end{mathletters}%
\vspace*{-.2in}
\beq
\delta \tilde{g}_{\mu\nu} = 0
\label{eq:1.16}
\eeq
The action (\ref{eq:1.13}) involves the anomaly density
${\cal A}(g) \propto \sqrt{-g} R(g)$,
and is the two-dimensional version of a general construction
\cite{ref:5}:
to obtain a given conformal anomaly (density) ${\cal A}(g)$
one defines a combination
$\tilde{g}_{\mu\nu} =  e^{-N\varphi} g_{\mu\nu}$,
which is Weyl-invariant for a selected value of $N$ with an
appropriate transformation law for $\varphi$, and takes as the
anomaly-generating action $I_{\cal A}$
\beq
I_{\cal A} \propto \int {\cal A}(\tilde{g}) \varphi
\label{eq:1.17}
\eeq
whose Weyl variation
(\ref{eq:1.15} -- \ref{eq:1.16})
requires varying only $\varphi$, thereby trivially
reproducing the anomaly.
\beq
\left.
{\delta I_{\cal A} \over \delta \varphi}
\right|_{\varphi=0} \propto {\cal A}(g)
\label{eq:1.18}
\eeq
For our case, ${\cal A}(g) = {1\over24\pi} \sqrt{-g} \, R(-g)$,
so that the integrand coincides with (\ref{eq:1.1}).
}

The system (\ref{eq:1.1}) implies the
scalar field equation (\ref{eq:1.3}),
while the metric variation of its action
[{\it c.f.} (\ref{eq:1.5})] is
\beq
T_{\mu\nu} \equiv
{2 \over \sqrt{-g}}
{\delta I \over \delta g^{\mu\nu}}
=
t_{\mu\nu} + \Delta_{\mu\nu}
\label{eq:1.19}
\eeq
\beq
t_{\mu\nu} =
-{1\over 48\pi} \left(
\partial_\mu \varphi \partial_\nu \varphi
- {1\over2} g_{\mu\nu} g^{\alpha\beta}
\partial_\alpha \varphi \partial_\beta \varphi \right)
{}~~,~~~~
\Delta_{\mu\nu} = -{1\over24\pi} (D_\mu D_\nu - g_{\mu\nu} D^2) \varphi ~~.
\label{eq:1.20}
\eeq
The total energy-momentum tensor $T_{\mu\nu}$ consists of the
conventional piece $t_{\mu\nu}$, and the improvement $\Delta_{\mu\nu}$
[which survives in the flat space,
Minkowski coordinate, limit as
\hbox{$-{1\over24\pi} (\partial_\mu
\partial_\nu - \eta_{\mu\nu} \Box) \varphi$}].
Covariant conservation of $T_{\mu\nu}$ is a consequence
of the action's coordinate invariance
and is guaranteed by (\ref{eq:1.3}).
[In flat space, $\Box \varphi$ vanishes and $T_{\mu\nu}$ is also traceless.]

As is well-known, in this one-component scalar field (rather than
string) context, treating the metric as a dynamical variable, thereby
setting $T_{\mu\nu}$ to zero, implies that $0 = T^\mu_\mu =
{1\over24\pi} D^2 \varphi$, and by (\ref{eq:1.3}), space is flat.
Consequently $\varphi$ is a free wave field
obeying $\Box \varphi = 0$.
Moreover, for a wave field,
$\varphi(x) = \varphi_+ (x^{+}) + \varphi_-(x^{-})$,
$x^{\pm} \equiv {1\over\sqrt{2}} (t \pm \sigma)$
and the remaining $T_{\mu\nu} = 0$ equations do not allow any
excitations in $\varphi$:
the field profiles are constrained to the forms
$\varphi_\pm (x^\pm) = 2 \ln \left( {x^\pm - x^\pm_0 \over a^\pm} \right)$,
where $x^\pm_0$ and $a^\pm$ are constants.
We remark that in this model, the
coupling of ``$\varphi$-matter'' to
``dynamical gravity'' ensures absence of both,
contrary to our usual ideas of gravitational
coupling and equivalence principle.
Note however, the negative sign of the action (\ref{eq:1.1}),
whose consequence is that $\varphi$-matter and gravity possess
kinetic terms with opposite signs,
and hence their dynamical effects cancel each other.

To avoid triviality, we regard gravity to be a background, and
$T_{\mu\nu}$ to be the scalar field's improved stress tensor, even if
space is flat (as we shall henceforth assume for simplicity).  In that case,
$\Delta_{\mu\nu}$ is a superpotential, identically conserved.  It plays
an essential role in the energy-momentum tensor algebra of the theory,
contributing to the Schwinger term or Virasoro central charge.

We recall that in any quantum theory with conventional positivity
properties, the equal-time commutator between energy and momentum
densities must acquire, in addition to the terms required by Poincar\'e
invariance, a non-vanishing central charge $c$ \cite{ref:6}.
In two dimensions, in particular, we must have
\beq
{}[T_{00} (\sigma), \, T_{01}(\tilde{\sigma})]
= i [T_{00}(\sigma) + T_{00}(\tilde{\sigma})]
\delta'(\sigma-\tilde{\sigma})
- ic \delta''' (\sigma-\tilde{\sigma})
\label{eq:1.21}
\eeq
where primes denote spatial derivatives
and the energy-momentum tensor components
are evaluated at a common time $t$, which is suppressed.
In addition to the quantal contributions to $c$
from normal-ordering the bilinears $t_{\mu\nu}$ \cite{ref:7},
there arise ``classical'' terms coming from $\Delta_{\mu\nu}$.
Using the definition (\ref{eq:1.20}),
and the fact that ${\partial \over \partial t} \varphi = \Pi$
is the conjugate momentum variable, we have
\beq
\Delta_{00} = -{1\over 24\pi} \varphi'' ~~,~~~~
\Delta_{01} = -{1\over 24\pi} \Pi'
\label{eq:1.22}
\eeq
and so
\beq
{}i [\Delta_{00} (\sigma), \Delta_{01} (\tilde{\sigma})]
= {1\over(24\pi)^2} \delta''' (\sigma-\tilde{\sigma}) ~~.
\label{eq:1.23}
\eeq
The remaining commutators in (\ref{eq:1.21}) are easily seen to reproduce
(only) the desired $T_{00}$ terms on the right side.  Hence the $WZ$ action
(\ref{eq:1.1}) has indeed, thanks to $\Delta_{\mu\nu}$, given rise to
the central charge in (\ref{eq:1.21}) in a simple free-field context
\cite{ref:8}.

\setcounter{section}{2}
\setcounter{equation}{0}

\bigskip

\noindent
{\bf II.}~
Is $\Delta_{\mu\nu}$ the only possible improvement term for our system?
In particular, can one define a chiral counterpart
$C_{\mu\nu} = C_{\nu\mu}$?
Even though there is no spin in two dimensions,
there still are right/left movers.
Let us first start with free fields in flat space.
Just as $\Delta^{\mu\nu}$ can be written as
\beq
\Delta^{\mu\nu}
= -{1\over24\pi}
\epsilon^{\mu\alpha}
\partial_\alpha
\epsilon^{\nu\beta}
\partial_\beta \varphi ~~,
\label{eq:2.1}
\eeq
an odd term involving only two derivatives is uniquely proportional to
\beq
C_{\mu\nu} \equiv -{1\over48\pi}
(\epsilon_{\mu}^{~\alpha} \partial_\alpha \partial_\nu +
\epsilon_\nu^{~\alpha} \partial_\alpha \partial_\mu) \varphi ~~.
\label{eq:2.2}
\eeq
(We use $\epsilon^{\alpha\beta} = -\epsilon^{\beta\alpha},
{}~\epsilon^{01} = 1$.)
Unlike $\Delta_{\mu\nu}$, $C_{\mu\nu}$ is conserved,
{\bf not} identically, but for wave fields:
$\partial_\mu C^{\mu\nu} = -{1\over48\pi} \epsilon^{\nu\alpha}
\partial_\alpha \Box \varphi$.
(It is, instead, identically traceless.)
This is of course sufficient for our purposes.  Let us calculate the
relevant components
$C_{00}, C_{01}$
\beq
C_{00} = -{1\over24\pi} \Pi' = \Delta_{01}
{}~~,~~~~
C_{01} = -{1\over24\pi} \varphi'' = \Delta_{00} \: .
\label{eq:2.3}
\eeq
These equalitites show the duality between the two improvement terms.

Before continuing our development, we must verify that $C_{\mu\nu}$ does
not affect the Poincar\'e generators $(P^\mu, M)$, {\it i.e.\/} that
\begin{mathletters}%
\label{eq:2.4}
\beq
\delta P^\mu = \int d\sigma \, C^{0\mu} ~~,~~~~
\delta M = \int d\sigma \, (\sigma C^{00} - t C^{01})
\label{eq:2.4a}
\eeq
vanish.  This is trivially verified for
$\delta P^\mu$
and for $\int d\sigma \, t \, C^{01}$
by (\ref{eq:2.3}), provided edge contributions can be dropped.
However, there remains
\beq
\delta M
= \int d\sigma \, \sigma \, C^{00}
= -{1\over24\pi} \int d\sigma \, \sigma \Pi'
\equiv {1\over24\pi} \int d\sigma \, \Pi
\label{eq:2.4b}
\eeq
\end{mathletters}%
which seems not to vanish.  To see that it does, we must use the wave
equation that ensures conservations of $C_{\mu \nu}$: any solution of
$\Box \phi = 0$ is the sum $\varphi_+ (x^{+}) + \varphi_-(x^{-})$,
so that
\hbox{$\Pi={\varphi_+}'-{\varphi_-}'$}
and this integrates to zero in
the absence of spatial edge terms.  (Being a superpotential, the
$\Delta_{\mu\nu}$ improvement automatically leaves the Poincar\'e
generators unaffected, absent end-point terms \cite{ref:9}.)

One may now consider the general improved energy-momentum tensor,
\beq
\overline{T}_{\mu\nu} \equiv t_{\mu\nu}
+ \alpha \Delta_{\mu\nu} + \beta C_{\mu\nu}
\label{eq:2.5}
\eeq
and evaluate (the $\Delta$- and $C$- dependent parts of) the commutator
(\ref{eq:1.21}).  Using (\ref{eq:2.3}) and paying attention to the oddness
of  $\delta'''$, we find that (\ref{eq:1.21}) is obeyed in terms of
$\overline{T}_{\mu\nu}$,
with $c = {1\over(24\pi)^2} (\alpha^2 + \beta^2)$, results that are indeed
exploited when
separately
calculating left/right contributions to the algebra in
more conventional approaches \cite{ref:1}.

Given the origin of the superpotential $\Delta_{\mu\nu}$ from a
nonminimal covariant coupling to gravity (a general property of true
superpotentials), one may ask whether $C_{\mu\nu}$ can likewise be
obtained from some original gravitational coupling.

Recall first the $\Delta_{\mu\nu}$ case:
we note the variational property
\begin{mathletters}%
\label{eq:2.6}
\beq
\delta (\sqrt{-g} R) =
{-\epsilon^{\mu\alpha} \epsilon^{\nu\beta} \over \sqrt{-g}}
\, D_\alpha D_\beta \, \delta g_{\mu\nu} ~~,
\label{eq:2.6a}
\eeq
so that
\beq
\delta \int d^2 x \sqrt{-g} \, R \varphi
= \int d^2 x \sqrt{-g} \, \delta g^{\mu\nu}
(D_\mu D_\nu - g_{\mu\nu} D^2) \varphi
\label{eq:2.6b}
\eeq
\end{mathletters}%
as was used for the derivation of $\Delta_{\mu\nu}$ in (\ref{eq:1.20}).
For the chiral version,
we need a quantity $C$
to multiply $\varphi$, with variation
\beq
\delta \int d^2 x \, C \varphi = \int d^2 x \, \delta g_{\mu\nu}
\epsilon^{\mu\alpha} D^\nu D_\alpha \varphi
\label{eq:2.7}
\eeq
which would give $C_{\mu\nu}$;
also the condition $D^2 \varphi = 0$
must be incorporated.
In fact a weaker requirement will suffice for our flat space
application: the variation (\ref{eq:2.7}) should hold in the
flat space limit.

We can immediately
state what this coupling {\it cannot\/} be: since $C_{\mu\nu}$ is
{\it not\/} identically conserved, it cannot
arise from the variation with respect to $g_{\mu\nu}$ of
an invariant action.
[Indeed, this is also clear because $\sqrt{-g} R$ is the only geometric
scalar density of second derivative order that can be used to multiply
$\varphi$ \cite{ref:10}.]


An imperfect solution to these requirements is constructed
in terms of the quantity
$R^\mu$, whose divergence is the Euler density.
\begin{eqnarray}
\partial_\mu R^\mu &=& \sqrt{-g} R
\label{eq:2.8}
\end{eqnarray}
Of course, $R^\mu$ is ambiguous up to terms of the form
$\epsilon^{\mu\nu} \partial_\nu r$, which would not contribute to
(\ref{eq:2.8}).  $R^\mu$ cannot be presented explicitly and locally in
terms of the generic metric $g_{\mu \nu}$ and its derivatives
$\partial_{\alpha} g_{\mu \nu}$ \cite{ref:11}; rather it is necessary to
parametrize $g_{\mu\nu}$.  Defining the unimodular metric
\beq
\gamma_{\mu\nu} = g_{\mu\nu} / \sqrt{-g} ~~,~~~~
\sqrt{-g} \equiv e^\sigma
\label{eq:2.9}
\eeq
and parametrizing $\gamma_{\mu\nu}$
and its inverse $\gamma^{\mu\nu}$,
by writing their light cone components as
\begin{mathletters}%
\label{eq:2.10}
\begin{eqnarray}
\gamma_{++} &=& - \gamma^{--} = e^\alpha \sinh \beta
\label{eq:2.10a} \\
\gamma_{--} &=& - \gamma^{++} = e^{-\alpha} \sinh \beta
\label{eq:2.10b} \\
\gamma_{+-} &=& \gamma_{-+} = \gamma^{+-} = \gamma^{-+} = \cosh\beta
\label{eq:2.10c}
\end{eqnarray}
\end{mathletters}%
gives for $R^\mu$
\beq
R^\mu = \gamma^{\mu\nu} \partial_\nu \sigma + \partial_\nu
\gamma^{\mu\nu} - \epsilon^{\mu\nu} (\cosh \beta - 1)
\partial_\nu \alpha \: .
\label{eq:2.11}
\eeq
The explicit parametrization (\ref{eq:2.10}) is needed to present
the last term in (\ref{eq:2.11}) \cite{ref:12}.

Even though the last contribution in (\ref{eq:2.11}) to $R^\mu$ is not
expressible in terms of $g_{\mu\nu}$ or $\gamma_{\mu\nu}$, its arbitrary
variation satisfies a formula involving only $\gamma_{\mu\nu}$

\beq
\delta \left[
\epsilon^{\mu\nu} (\cosh \beta - 1) \partial_\nu \alpha
\right]
- \partial_\nu
\left[
\epsilon^{\mu\nu} (\cosh \beta - 1) \delta\alpha
\right]
= -{\textstyle{1\over2}} \gamma^{\mu\nu}
\left(
\partial_\alpha \gamma_{\nu\beta}
+ \partial_\beta \gamma_{\nu\alpha}
- \partial_\nu \gamma_{\alpha\beta}
\right)
\delta \gamma^{\alpha\beta} \: .
\label{eq:2.12}
\eeq
The second term on the left is a curl and so will not contribute;
the right side may be written as
$-\gamma^\mu_{\alpha\beta} \delta \gamma^{\alpha\beta}$,
where $\gamma^\mu_{\alpha\beta}$ is the Christoffel affinity of the
$\gamma_{\mu\nu}$ metric: $\gamma_{\alpha\beta}^\mu =
\Gamma_{\alpha\beta}^\mu \Bigr|_{g_{\mu\nu} = \gamma_{\mu\nu}}$.

We remark that the last term in (\ref{eq:2.11})
naturally defines a 1-form
$a \equiv (\cosh\beta - 1) d\alpha$ and the 2-form
$\omega = da = \sinh\beta \, d\beta \, d\alpha$.
These are recognized as the canonical 1-form and the symplectic 2-form,
respectively, for SL(2,$R$).  Indeed $\omega$ also equals
${1\over2} \epsilon_{abc} \xi^{a} d \xi^{b} d \xi^{c}$,
where $\xi^a$ is a three-vector on a hyperboloid = SL (2,$R$)/U(1) :
$(\xi^1)^2 - (\xi^2)^2 - (\xi^3)^2 = -1$.
Effectively, $\omega$ is the Kirillov-Kostant 2-form on SL(2,$R$)
\cite{ref:13}.

While the divergence of $R^\mu$ is the scalar curvature density,
$R^\mu$ itself is not a vector density under coordinate transformations.
Rather for the infinitesimal diffeomorphism (\ref{eq:1.8})
one verifies that there is an additional, identically conserved,
non-tensor term in the transformation law
\begin{mathletters}%
\label{eq:2.13}
\beq
\delta_f (R^\mu / \sqrt{-g}) = L_f (R^\mu / \sqrt{-g}) +
\epsilon^{\mu\nu} \partial_\nu \Delta_f / {\sqrt{-g}}
\label{eq:2.13a}
\eeq
\beq
\Delta_f \equiv
(\partial_+ - e^\alpha \tanh {\beta\over 2} \partial_- )
f^+
-
(\partial_- - e^{-\alpha} \tanh {\beta\over 2} \partial_+)
f^- \: .
\label{eq:2.13b}
\eeq
\end{mathletters}%
Consequently, a world scalar action
equivalent to (\ref{eq:2.6b})
may be constructed by coupling
vectorially $R^\mu$ to a scalar field $\varphi$,
$I_V = \int d^2 x \, R^\mu \, \partial_\mu \varphi$.
[The formula (\ref{eq:2.6b}) is preferred in that no derivatives act on
$\varphi$ and the symplectic structure is preserved.]
But now we see that the axial action
\beq
I_C = \int d^2 x {\epsilon^{\mu\nu} \over \sqrt{-g}}
R_\mu \partial_\nu \varphi
=
\int d^2 x \, \varphi \, \epsilon^{\mu\nu} \, \partial_\mu
\, \left( {R_\nu \over \sqrt{-g}} \right)
\label{eq:2.14}
\eeq
will be insensitive both to the variation (\ref{eq:2.13}),
and to the ambiguity $\epsilon^{\mu\nu} \partial_{\nu} r$
in $R^\mu$ if $\varphi$ satisfies $D^2 \varphi = 0$
[The last form of (\ref{eq:2.14}) preserves the symplectic structure of
the $\varphi$ field.]
Thus we attempt generating $C_{\mu\nu}$ from $I_C$.


When $I_C$ is varied, one finds using (\ref{eq:2.12}) that
\beq
\delta I_C = \int \! d^2 x \, \delta g_{\mu\nu}
\left[
\epsilon^{\mu\alpha}
\left( \! D^\nu \! + {R^\nu \over \sqrt{-g}} \right)
D_\alpha \varphi - g^{\mu\nu} {R_\alpha \over \sqrt{-g}}
\epsilon^{\alpha\beta} \partial_\beta \varphi
\right]
- \int \! d^2 x \, \sqrt{-g} \, \delta \alpha
\left[ \cosh \beta - 1 \right]
D^2 \varphi
\label{eq:2.15}
\eeq
The second integral may be dropped when $D^2 \varphi = 0$, and
the desired result (\ref{eq:2.7}) is achieved, in flat
space where $R^\mu$ vanishes.  Thus the chiral improvement
(\ref{eq:2.2}) may be derived by varying ${1\over 48\pi} I_C$, and
passing to flat space.
But what remains unsatisfactory about the $C_{\mu\nu}$ construction,
as compared to that of $\Delta_{\mu\nu}$, is our inability
to present
a single action of the form $I[\varphi] + I_C$, such that its
$\varphi$-variation implies the required $D^2 \varphi = 0$ and flatness
conditions.

The explicit tracelessness of $C_{\mu\nu}$ is now also understood:
from (\ref{eq:2.11}) and (\ref{eq:2.14}) we see that $I_C$ does not
depend on the conformal factor $\sigma$, and is therefore
Weyl invariant, even when it is not diffeomorphism invariant for
$D^2 \varphi \neq 0$.

We conclude that while the two improvements we have discussed
are both permitted (and useful) additions to the flat space scalar field's
stress tensor, they have quite different status
when related to
underlying gravitational couplings.


\begin{thebibliography}{99}


\bigskip
\frenchspacing


\bibitem{ref:1}
For a review, see for example D.~Friedan, E.~Martinec and
S.~Shenker, {\it Nucl.~Phys.\/}~{\bf B271}, 93 (1986).

\bibitem{ref:2}
A.~Polyakov, {\it Mod.~Phys.~Lett.\/}~{\bf A2}, 893 (1987);
{\it Gauge Fields and Strings\/} (Harwood, New York, NY, 1987);
see also M.~Duff, {\it Nucl.~Phys.\/}~{\bf B125}, 334 (1977).

\bibitem{ref:3}
This alternative determination of $\theta_{\mu\nu}$ was given by M.~Bos,
{\it Phys.~Rev.~D\/}~{\bf 34}, 3750 (1986).

\bibitem{ref:4}
It is interesting to note that alternative
procedures for handling infinities are available and that it is possible to
maintain Weyl invariance, at the expense of diffeomorphism invariance.
Nevertheless, diffeomorphisms with unit Jacobian, {\it i.e.}, in (1.8)
$\partial_\mu f^\mu = 0$, are maintained.  With this approach, the
energy-momentum tensor remains traceless, but its divergence becomes
$-{1\over 48\pi} \partial_\mu R$,
a formula whose structure is in fact dictated by the restricted
diffeomorphism symmetry.
``Physical'' results, like two-dimensional Hawking radiation,
remain unaffected.  For details see: D.~Karakhanyan, R.~Manvelyan and
R.~Mkrtchyan, {\it Phys.~Lett.\/}~{\bf B329}, 185 (1994);
G.~Amelino-Camelia, D.~Bak and D.~Seminara,
{\it Phys.~Lett.\/}~{\bf B354}, 213 (1995);
J.~Navarro-Salas, M.~Navarro and C.~Talavera,
{\it Phys.~Lett.\/}~{\bf B356}, 217 (1995);
and G.~Amelino-Camelia and D.~Seminara,
MIT preprint MIT--CTP--2443.

\bibitem{ref:5}
S.~Deser and A.~Schwimmer, in preparation.

\bibitem{ref:6}
D.~Boulware and S.~Deser, {\it J.~Math.~Phys.\/}~{\bf 8}, 1468 (1967).

\bibitem{ref:7}
In the two-dimensional case, these are finite and were calculated long ago by
S.~Fubini, A.~Hanson and R.~Jackiw,
{\it Phys.~Rev.~D\/}~{\bf 7}, 1732 (1972)
as well as by S.~Ferrara, R.~Gatto and A.~Grillo,
{\it Nuevo Cim.\/}~{\bf 12A}, 959 (1972).

\bibitem{ref:8}
That the improved energy-momentum tensor possess a {\it canonical\/}
triple-derivative Schwinger term in its equal time commutator algebra was
noted by C.~Callan, S.~Coleman and R.~Jackiw,
{\it Ann.~Phys.\/} (NY) {\bf 59}, 42 (1970).

\bibitem{ref:9}
The improvement does modify the dilation and special conformal
generators in dimensions higher than two, and for the Liouville theory
in two dimensions.  In these instances, the unimproved energy-momentum
tensor is not traceless, so that dilation and special conformal
generators cannot be given by moments of this energy-momentum tensor,
see {\it e.g.\/} Ref.~[8] and Jackiw in {\it Progress in Quantum Field
Theory\/}, H.~Ezawa and S.~Kamefuchi, eds.  (North-Holland, Amsterdam,
Netherlands 1986).

\bibitem{ref:10}
R.~Jackiw, C.~Teitelboim in
{\it Quantum Theory of Gravity\/}, S.~Christensen, ed.
(Adam Hilger, Bristol, UK, 1984).

\bibitem{ref:11}
In the {\it Zweibein\/} $(e_\alpha^a)$-connection
$(\omega_\mu \equiv \omega_{\mu a b} \epsilon^{ab})$
form of the Euler density
$\sqrt{-g} R =  \epsilon^{\mu\nu} \partial_\mu \omega_{\nu}$,
$R^\mu$ {\it is\/} explicitly defined
(up to the ambiguity $\epsilon^{\mu\nu} \partial_\nu r$)
without parametrization.
However, $R^\mu$ now no longer depends solely on the metric, but involves the
{\it Zweibein\/} in an essential way, since the connection is
$\omega_\nu =
{\epsilon^{\alpha\beta} \over \sqrt{-g}} \partial_\beta g_{\alpha\nu}
+ \epsilon_{ab} g^{\alpha\beta} e_\alpha^a \partial_\nu e_\beta^b$.
The first term is metric, but the second one clearly depends on the
essentially non-metric part of the {\it Zweibein\/}.

\bibitem{ref:12}
This is analogous to what happens with a Chern-Simons term.  Upon
performing a gauge transformation with a gauge function $U$, the
Chern-Simons term changes by a total derivative.  However, direct
evaluation of the gauge response includes the expression
$\omega = {1\over 24 \pi^2} {\rm ~tr~} \epsilon^{\alpha\beta\gamma}
U^{-1} \,
\partial_\alpha  U U^{-1}
\partial_\beta     U^{-1}
\partial_\gamma  U$,
which can be recognized as a total derivative only after $U$ is
explicitly parametrized.  For example, in SU(2),
$U = \exp \theta$, $\theta = \theta^a \sigma^a / 2i$,
and $\omega = \partial_\alpha \omega^\alpha$ where
$\omega^\alpha = {1\over 4 \pi^2} {\rm ~tr~} \epsilon^{\alpha\beta\gamma}
\theta \partial_\beta \theta \partial_\gamma \theta
\left( {|\theta| - \sin|\theta| \over |\theta|^3} \right)$
with $|\theta| \equiv \sqrt{\theta^a \theta^a}$; see Jackiw in
S.~Treiman, R.~Jackiw, B.~Zumino and E.~Witten,
{\it Current Algebra and Anomalies\/},
(Princeton University Press / World Scientific, Princeton, NJ /
Singapore, 1985).

\bibitem{ref:13}
We thank V.~P.~Nair for pointing this out.

\nonfrenchspacing
\end{thebibliography}
\end{document}